\documentclass{ws-procs975x65}

\begin{document}

\title{RELATIVISTIC SPIN-PRECESSION IN BINARY PULSARS}

\author{MICHAEL KRAMER$^*$}

\address{Max-Planck-Institut f\"ur Radioastronomie\\
Auf dem H\"ugel 69\\
53121 Bonn, Germany\\
and Jodrell Bank Centre for Astrophysics, University of Manchester\\
Turing Building, Oxford Road \\
Manchester M13 9PL, UK \\
$^*$E-mail: mkramer@mpifr.de}

\begin{abstract}
  After the first prediction to expect geodetic precession in binary
  pulsars in 1974, made immediately after the discovery of a pulsar with a
  companion, the effects of relativistic spin precession have now been
  detected in all binary systems where the magnitude of the precession
  rate is expected to be sufficiently high. Moreover, the first
  quantitative test leads to the only available constraints for
  spin-orbit coupling of a strongly self-gravitating body for general
  relativity (GR) and alternative theories of gravity. The current
  results are consistent with the predictions of GR, proving the
  effacement principle of spinning bodies. Beyond tests of theories of
  gravity, relativistic spin precession has also become a useful tool
  to perform beam tomography of the pulsar emission beam, allowing to
  infer the unknown beam structure, and to probe the physics of the
  core collapse of massive stars.
\end{abstract}


\bodymatter

\section{Introduction}\label{aba:sec1}

The theory of general relativity predicts, as other alternative
relativistic theories of gravity, that space-time is curved by the
presence of matter. It is this curvature that then describes how
matter is moving. A direct way to verify this concept of curved
space-time is to measure, for instance, a deflection of
light\cite{wil06} or a ``Shapiro-delay'' \cite{sha64}, i.e.~an extra
flight-time of an electromagnetic signal when it passes near a mass
and transverses its curved space-time compared to the flight-time in
flat space-time away from masses (e.g.~Ref.~\citen{ksm+06}). Another
way of measuring the impact of curved space-time is the usage of a
gyroscope orbiting a central mass. This effect, known as geodetic
precession or de Sitter precession represents the effect on a vector
carried along with an orbiting body such that the vector points in a
different direction from its starting point (relative to a distant
observer) after a full orbit around the central
object\cite{des16,fok21}. Experimental verification has been achieved
by precision tests in the solar system, e.g. by Lunar Laser Ranging
(LLR) measurements, or recently by measurements with the Gravity
Probe-B satellite mission (see Ref.~\citen{wil06} for a review of
experimental tests). However, these tests are done in the weak field
conditions of the solar systems, while it is important to also perform
tests in the strong-field regime. What is therefore needed are compact,
spinning test masses where we can infer and monitor the orientation of
the spin direction when they are in motion about a binary
orbit. Fortunately, nature provides us with such ideal labs: binary
radio pulsars.

In binary system one can interprete the observations, depending on the
reference frame, as a mixture of different contributions to
relativistic spin-orbit interaction. One contribution comes from the
motion of the first body around the centre of mass of the system
(deSitter-Fokker precession), while the other comes from the dragging
of the internal frame at the first body due to the translational
motion of the companion \cite{ber75}. 
Hence, even though we loosely talk about
geodetic precession, the result of the spin-orbit coupling for binary
pulsar is more general, and hence we will call it {\em relativistic
  spin-precession}.

The consequence of relativistic spin-precession is a precession of
the pulsar spin about the total angular moment vector, changing the
orientation of  the pulsar relative to Earth. This allows us to detect
this effect, to study its impact on the observed pulsar emission, to
gain information about the pulsar itself and, last but not least, to
test the predictions for relativistic spin-orbit coupling in general
relativity and alternative theories of gravity.

This contribution will briefly review the basic properties of pulsars that
are needed to understand and interpret the seen effects in pulsars,
before it presents a summary of those observations where relativistic
spin-orbit coupling has been detected and studied. Finally, I will
summarize the applications of this effect in fields ranging from core 
collapse physics in massive stars to testing theories of gravity.

\section{Pulsars}

Pulsars are rotating neutron stars that emit a radio beam that is
eventually powered by the pulsars' rotational energy and that is
centred on the magnetic axis of the neutron star. As the magnetic axis
and the hence the beam are inclined to the rotation axis, the pulsar
acts as a cosmic lighthouse, and a pulsar appears as a pulsating radio
source. The moment of inertia and the stored rotational energy of
pulsars are large, so that in particular the fast rotating millisecond
pulsars deliver a radio ``tick'' per rotation with a precision that
rivals the best atomic clocks on Earth. Corresponding pulse (or spin)
periods range from 1.4 ms to 8.5 s. As they concentrate an average of
1.4 solar masses on a diameter of only about 20 km, pulsars are
exceedingly dense and compact, representing the densest matter in the
observable universe. The resulting gravitational field near the
surface is large, enabling strong-field tests of gravity. Here we
describe a cosmic experiment where we see these spinning tops orbiting
another massive companion. In order to infer and monitor their spin
precession, we do not utilize the accurate time-of-arrival
measurements of their pulses as done in most other pulsar timing
experiments, but we study the pulsar radio emission.

\subsection{Pulsar radio emission}
 
\begin{figure}[t]
\begin{center}
\psfig{file=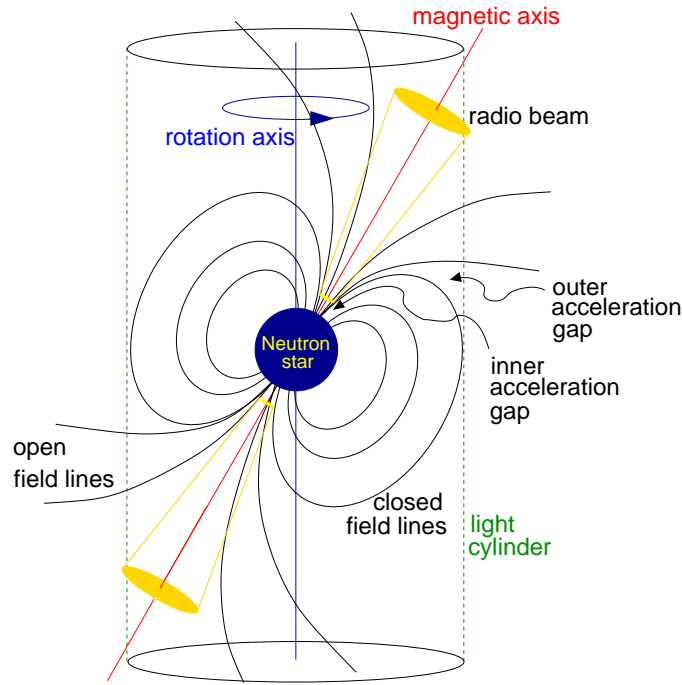,width=10cm}
\end{center}
\caption{Basic model of a radio pulsar. Taken from Ref.~\citen{lk05}. }
\label{fig:pulsar}
\end{figure}

After more than 40 years of pulsar research, the details of the actual
emission process still elude us, but our basic understanding is
sufficient to perform the experiments described later.  In our
straw-man model, the high magnetic field of the rotating neutron star
($B_{\rm surf} \sim 10^8 - 10^{14}$ Gauss) induces a huge electric
quadrupole field and an electromagnetic force that exceeds gravity by
ten to twelve orders of magnitudes. Charges are pulled out easily from
the surface, and the result is a dense, magnetized plasma that
surrounds the pulsar. The strong magnetic field forces the plasma to
co-rotate with the pulsar like a rigid body. This co-rotating {\em
  magnetosphere} can only extend up to a distance where the
co-rotation velocity reaches the speed of light\footnote{Strictly
  speaking, the Alfv\'en velocity will determine the co-rotational
  properties of the magnetosphere.}. This distance defines the
so-called light cylinder which separates the magnetic field lines into
two distinct groups, i.e.~{\em open and closed field lines}. Closed
field lines are those which close within the light cylinder, while
open field lines would close outside. The plasma on the closed field
lines is trapped and will co-rotate with the pulsar forever. In
contrast, plasma on the open field lines can reach highly relativistic
velocities and can leave the magnetosphere, creating the observed
radio beam at a distance of a few tens to hundreds of km above the
pulsar surface (see Fig.~\ref{fig:pulsar}).

\begin{figure}[t]
\begin{center}
\psfig{file=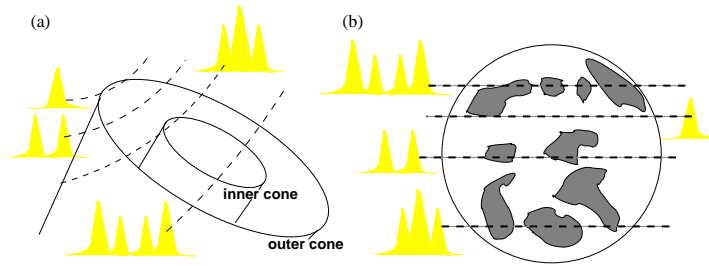,angle=-90,width=10cm}
\end{center}
\caption{Creation of multi-component pulse profiles in two possible beam
models. (a) A nested cone structure,  (b) a patchy beam structure. See
text for details.  Taken from Ref.~\citen{lk05}. }
\label{fig:beam}
\end{figure}

\subsubsection{The structure of the pulsar beam}

The radio beams show high time-variability in individual recorded
pulses that reflects the instantaneous plasma processes in the pulsar
magnetosphere at the moment when the beam is directed towards Earth.
Despite this variety displayed by the single pulses, the mean pulse
shape computed by averaging a few hundred to few thousand of pulses
is usually very stable\cite{lk05}.  In contrast to the snapshot
provided by the individual pulses, the average pulse shape, or {\em
  pulse profile}, can be considered as a long-exposure picture,
revealing the global circumstances in the magnetosphere. These are
mostly determined by geometrical factors and the strong magnetic
field.  Apart from a distinct evolution with radio frequency, the same
profiles are obtained, no matter where and when the pulses used to
compute the average have been observed. Changes in the pulse profile
therefore usually reflect a change in the pulsar geometry that can be
detected and monitored. 
 \footnote{ For completeness we note that Lyne et al. have found
evidence for long-tern changes in the pulse profile which are
correlated with changes in the pulse-spin down.}

The observed pulse profiles show a large variety of shapes. Although
each pulsar exhibits a slightly different profile -- almost like a
unique fingerprint -- a systematic pattern can be recognized. The most
simple model successfully describing the beam shapes is that of a
hollow cone of emission\cite{kom70}. It is based on the idea that the
outermost open field lines, which show the largest curvature among the
``emitting field lines'', should be associated with the strongest
emission, leading naturally to a cone-like structure. Observations
show that this picture is vastly oversimplified, since we also often
observe components inside this hollow-cone structure. Those inner
component may be positioned in preferred located (e.g. in the cone
centre as a ``core'' component \cite{ran90} and/or in a nested cone
structure \cite{ran93}) or in a seemingly random pattern of a patchy
beam shape \cite{lm88} (see Fig.~\ref{fig:beam}). It is clear, however, that the observed pulse
profile depends eventually on how our line-of-sight cuts the
emission cone. With relativistic spin precession changing our path
through the beam, we therefore not only expect the observed pulse
shape to change, but we also have the chance to actually probe the beam
structure in latitudinal direction to test specific emission theories
(see Section~\ref{sec:app:beam}).

\subsubsection{Polarisation --- Signatures of Geometry}

An important and most useful property of pulsar radio emission is its
typical high degree of polarisation. The radiation is often 100\%
elliptically polarized, and it is usual practice to separate the
polarisation into linearly and circularly polarized components. The
linear component is often the far dominating one, although pulsars
with circular components as strong as 30\% or more are not uncommon.
Polarisation serves as a useful diagnostic tool to obtain information
about the viewing geometry \cite{kj08}.

The strong coupling of the outwards-moving plasma to the magnetic
field lines in the pulsar magnetosphere has the effect that the plane
of polarisation of the linear component is determined by the plane
embedding the corresponding field line. The observed position angle
(PA) of the linear polarisation is then given by the projection of
this direction onto our line-of-sight. The result is an S-like curve
of the PA whose shape depends on the angle between the rotation and
magnetic axes, $\alpha$, as well as on the distance of our
line-of-sight to the magnetic pole, $\beta$ (see inset in
Fig.~\ref{fig:1906}). If our line-of-sight cuts the emission beam
close to the magnetic axis, the PA changes rapidly when crossing the
centre of the beam. If the impact angle $\beta$ is large and we are
cutting the cone further away from the pole, the transition is much
smoother and the PA swing much flatter \cite{lk05}.

By measuring the polarisation characteristics of a pulsar, we can in
principle win information about the pulsar's orientation towards
us. In practice, fitting this {\em rotating vector model}\cite{rc69a}
(RVM) turns out to be often difficult. Although the majority of
observed PA swings can be well described by the RVM after correcting
for sometimes occurring orthogonal modes (i.e.~jumps of the PA by
nearly 90$^\circ$ which are probably magnetospheric propagation
effects), the uncertainties in the obtained angles representing the
geometry are typically large. The reason is not a failure of the
model, but simply the small size of beam of most pulsars, which
provides constraints to the fit for only the small fraction of the
pulse period when the pulse is actually observed, which is typically
only 4\% (see Ref.~\citen{lk05} for more details).

Observations of normal (i.e.~non-recycled pulsars, see below for the
evolutionary
differences) pulsars
confirm the geometrical meaning of the RVM \cite{kj08}. This is less
clear for recycled pulsars (e.g.~Ref.~\citen{xkj+98}),
but detecting a change in the observed PA swing curve will immediately
indicate a change in viewing geometry. Another way to detect changes
in the viewing geometry was pointed out recently by Kramer \& Wex
(2009, Ref.~\citen{kw09}) who presented a method to use the absolute position angle
measurement (i.e.~a PA measurement tied to a celestial reference
frame) to determine a change in the spin direction of pulsars. This
method has now been successfully applied to PSRs J1141$-$6545 and
J1906$+$0746 (see Fig.~\ref{fig:1906}) as described later.

\subsubsection{Formation \& Evolution}
\label{sec:evolv}

Relativistic spin precession in binary pulsars only occurs if the
pulsar spin vector is misaligned with the total angular momentum
vector. If precession is observed, observations can be used to
determine this misalignment angle. The question as to whether the spin
vector is aligned or not depends on the evolutionary history of the
pulsar and the binary system as a whole.

Neutron stars and pulsars are born in a supernova explosions,
conventionally believed to be the core collapse of a massive star
although alternative routes seem to exist (see Section
\ref{sec:app:sn}). Most pulsars loose a possible companion in this
supernova explosion but those which retain it may be spun-up in a
later period of their life. Those pulsars appear as millisecond (or
recycled)  pulsars with short spin period and are expected to be still
in orbit with the remnant of the companion star. 

The mass of the
companion star determines the duration of the accretion process and
the fate of the system. If the companion is of low mass, it evolves
slowly and the mass transfer can be sufficiently long to spin up the
pulsar to periods of a few milliseconds or less. The end product would
be a pulsar - white dwarf system where due to the exchange of angular
momentum and tidal interactions all spin vectors are expected to be
aligned, so that no spin-precession should occur.

If the companion star is massive enough to undergo a supernova
explosion on its own, the evolution timescale is shorter and the
accretion process is cut short. The result for the pulsar
is a spin period of tens of millisecond and a neutron star companion,
should the system survive this second supernova explosion. The
formation of the second-born neutron star is likely to occur in an
{\em asymmetric supernova explosion} which imparts a kick on the
companion that tilts the new orbit relative to the pre-supernova
configuration. The spin vector of the recycled pulsar is unaffected
and now misaligned with the new total angular momentum vector --
relativistic spin procession should occur. Variations to the latter statement
are possible if the second supernova was ``gentle'' and produced only
a low-velocity kick to the second neutron star. In this case, as it
seems to be observed in the Double Pulsar (see below), the orbital
plane direction is hardly changed and the recycled pulsar may not
precess notably.

Under certain conditions, it is possible that we see radio emission of
the young, second-born pulsar in a binary system. This seems to be the
case in J1141$-$6545, J1906+0646 and, of course, in the Double Pulsar.
Here, we have no constraints on the possible direction of the pulsar
spin vector as this is determined by the individual properties of the
corresponding supernova explosion. It is therefore highly likely that
spin vectors are misaligned, so that spin precession is expected to be
observed.

\section{The experiments}

Since the orbital angular momentum is much larger than the angular
momentum of the pulsar, the orbital spin practically represents a
fixed direction in space, defined by the orbital plane of the binary
system. Therefore, if the spin vector of the pulsar is misaligned with the
orbital spin, relativistic spin-precession leads to a change in
viewing geometry, as the pulsar spin precesses about the total angular
momentum vector.  Consequently, as many of the observed pulsar
properties are determined by the relative orientation of the pulsar
axes towards the distant observer on Earth, we should expect a
modulation in the measured pulse profile properties, namely its shape
and polarisation characteristics. This was immediately recognized in
the ground-braking work by Damour \& Ruffini (Ref.~\citen{dr74}). Shortly
after the discovery of the first binary pulsar by Hulse \& Taylor in
1974 -- and even before its publication \cite{ht75a} -- Damour \&
Ruffini pointed out that such a modulation should occur with a
periodicity of that of the precession period.

The precession rate as predicted by general relativity (GR) 
is given by\cite{bo75b,ber75}
\begin{equation}
\label{eqn:om}
\Omega_{\rm p}  =  T_\odot^{2/3} \times \left( \frac{2\pi}{P_{\rm b}}\right)^{5/3} \times
 \frac{m_{\rm c}(4m_{\rm p}+3m_{\rm c})}{2(m_{\rm p}+m_{\rm c})^{4/3}} \times
 \frac{1}{1-e^2}
\end{equation}
where $P_{\rm b}$ is the period and $e$ the eccentricity of the orbit. We express
the masses $m_{\rm p}$ and $m_{\rm c}$ in units of solar masses ($M_\odot$) and the define
the constant $T_\odot=GM_\odot/c^3=4.925490947 \mu$s. $G$ denotes the
Newtonian constant of gravity and $c$ the speed of light. 
It is useful to note that, in GR, for equal masses
\begin{equation}
\Omega_{\rm p} = \frac{7}{24}\;\dot{\omega}
\sim 0.3\;\dot{\omega}.
\end{equation}

Spin precession may have a direct effect on the timing, as it causes
the polar angles of the spin and, hence, the aberration parameters of
the timing model to change with time (see Refs.~\citen{dt92,lk05} for
details). However, the consequences for the observed emission
properties are, usually, much more apparent and easier to identify as changes
in the timing parameters may get absorbed into other parameters for
a limited observing span.

Due to the changing cuts through the emission beam as the pulsar spin
axis precesses, we firstly expect the profile to narrow or widen
depending on the precession phase and beam structure. As this also
changes the distance of the observer's line-of-sight to the magnetic
axis, the position angle swing should become flatter or steeper,
depending on whether the impact angle $\beta$ is decreasing or increasing,
respectively. In comparison to the observed total power profile,
we can also expect the polarisation properties to be more
sensitive to the local conditions in the magnetosphere that are probed
by our line-of-sight, suggesting that changes in polarised emission go
further than just changing the position angle slope.

In order to see a measurable effect in any binary pulsar, {\em a)} the spin
axis of the pulsar needs to be misaligned with the total angular
momentum vector and {\em b)} the precession rate must be sufficiently
large compared to the available observing time to detect a change in
the emission properties. Table 1 lists the known Double
Neutron Star Systems (DNS) which typically show the largest degree of
relativistic effects due to the often short eccentric binary
orbits. However, the last entry in the table is PSR J1141$-$6545 which
is a relativistic system with a white dwarf companion \cite{acw+10}. 

Those pulsars that are marked with an asterisk have been identified as
pulsars showing relativistic spin precession. We will discuss them in
detail later below. What is apparent by inspecting the precession rate
as computed by Eqn.~(\ref{eqn:om}) is that the top 5 out of 8 sources
with a value for the expected precession rate indeed show the
effect. The only exceptions are pulsar A in the Double Pulsar
(i.e.~PSR J0737$-$3039) and PSRs J1756$-$2251 and J1829+2456. We
understand the lack in pulsar A by a possible alignment of its spin
axis with the orbital spin for which we have independent evidence that
will be discussed in Sec.~\ref{sec:app:sn}. PSR J1756$-$2251 is a
relatively new discovery \cite{fsk+04} with a rather simple profile,
so that a detection has not been made yet. The expected detection rate
for J1828+2456 is very likely too small to lead to significant
observable effects. In contrast, PSR B1534+12 has a relatively small
precession rate of 0.5 deg yr$^{-1}$ but as the second DNS discovered
it is well studied with a long history of excellent polarisation data
\cite{sta04}. Indeed, a third criterion {\em c)} that has been usually
important for actually making a firm detection of spin precession is
the usage or availability of long-term observations with identical or
similar instrumental set-up. Nowadays, as coherent de-dispersion
systems with many-bit digitization deliver nearly identical results,
observations made at different telescopes or with different data
acquisition systems can be combined much easier.

\begin{table}
\label{tab1}
\tbl{DNSs sorted according to the expected relativistic spin precession rate.
  Also included is PSR   J1141$-$6545 which is in a 
  relativistic orbit about a white dwarf companion. Pulsars marked
  with an asterisk have been identified of showing spin precession.
 For sources where no precession rate is listed,
  the companion mass could not be accurately measured yet, indicating 
  however, that the precession rate is low.}
{\begin{tabular}{lrcccc}
    \hline
    \hline
    PSR & $P$(ms) & $P_{\rm b}$ (d) &  $x$(lt-s) & $e$& $\Omega_{\rm p}$ (deg yr$^{-1}$) \\
    \hline
    J0737$-$3039A/B$^\ast$ & 
    \multicolumn{1}{c}{22.7/2770} & 0.10 & 1.42/1.51 & 0.09 & \multicolumn{1}{c}{4.8/5.1} \\
    J1906+0746$^\ast$ & 144.1 & 0.17 & 1.42 & 0.09 & 2.2 \\
    B2127+11C$^\ast$ & 30.5 & 0.34 & 2.52 & 0.68 & 1.9 \\
   B1913+16$^\ast$ & 59.0 & 0.33 & 2.34 & 0.62 & 1.2 \\
    J1756$-$2251 & 28.5 & 0.32 & 2.76 & 0.18 & 0.8 \\
    B1534+12$^\ast$ & 37.9 & 0.42 & 3.73 & 0.27	& 0.5 \\
   J1829+2456 & 41.0 & 1.18 & 7.24 & 0.14 & 0.08 \\
  \noalign{\smallskip}
    J1518+4904 & 40.9 & 8.64 & 20.0 & 0.25 & -- \\
   J1753$-$2240 & 95.1 & 13.63 & 18.1 & 0.30 & -- \\
   J1811$-$1736 & 104.2 & 18.8 & 34.8 & 0.83 & -- \\
  \noalign{\smallskip}
   J1141$-$6545$^\ast$ & 394.0 & 0.20 & 1.89 & 0.17 & 1.4 \\
    \hline
\end{tabular}}
\end{table}

\subsection{PSR B1913+16 -- The Hulse-Taylor Pulsar}

\label{sec:1913}

The first binary pulsars discovered \cite{ht75a} was also the first
DNS to be found, and the system remained the most relativistic one known
until the discovery of the Double Pulsar (see Sec.~\ref{sec:0737}). With its
discovery it was immediately recognized that it would be a superb
laboratory for relativistic gravity \cite{ht75a}. In particular the possible
detection of spin precession was pointed out from the start\cite{dr74}. The system
parameters in Table~1 infer a precession rate that is
relatively high, indeed promising to detect the effect on a reasonable
time scale. Weisberg et al.~(1989, Ref.~\citen{wrt89}) showed clearly that in
about 10 years of observations, the two prominent peaks in the pulse
profile had changed their relative amplitude significantly by about
1.2\% per year. While this change was attributed to the
effects of spin precession, this study did not detect the expected change
in pulse width, unfortunately. A polarisation study \cite{cwb90} was also
inconclusive as that changes in the polarisation properties that could
be attributed to spin precession were not found. The authors pointed
out that the beam pattern could be patchy or that the presence of a
``core'' beam component could disturb the regular S-like
like position angle swing that one would expect to see changing.

\begin{figure}[t]
\begin{center}
\begin{tabular}{lr}
\psfig{file=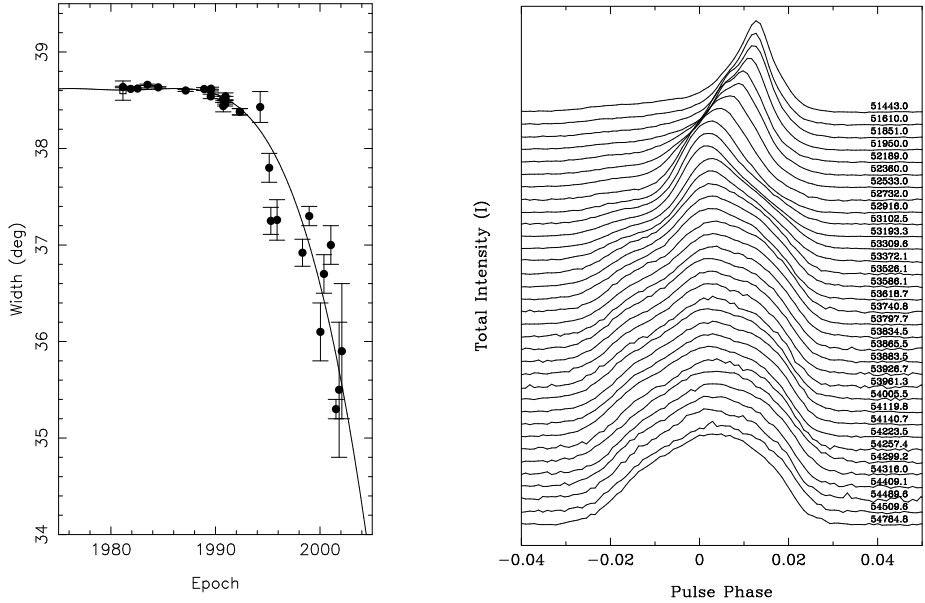,width=5.5cm} \hspace{0.5cm} & \psfig{file=fig3b.ps,width=6cm} 
\end{tabular}
\end{center}
\caption{Pulse changes caused by relativistic spin precession.
Left) Observed component separation for PSR B1913+16 (Kramer,
  priv. comm.), Right) measured
pulse profiles for PSR J1141$-$6545 \cite{mks+10}. The right hand
scale gives the observing date in Modified Julian Day, spanning from
Sep 22, 1999, at the top to Nov 11, 2008, at the bottom.}
\label{fig:1913}
\label{fig:1141}
\end{figure}

In 1998, using data from the Effelsberg telescope, combined with the original
Arecibo observations by Taylor, Weisberg and co-workers, it was
possible to show that the change in relative amplitude was continuing
with a rate found by Weisberg et al.~\cite{kra98}. Moreover, for the
first time a narrowing of the pulse profile was detected, in
accordance with a simple cone-like beam model (see
Fig.~\ref{fig:1913}). 
Fitting such a model to
the observed profile width data, the full geometry of the system
 could be determined, resulting in a measurement of the
angle between the pulsar spin and the total angular momentum vector,
which is an imprint of the asymmetric supernova explosion forming
this system (see Section~\ref{sec:app:sn}). The fit to the data also
predicts that the pulsar will disappear by moving out of our
line-of-sight around the year 2025.
These results were later confirmed by Weisberg \& Taylor
(Refs.~\citen{wt00,wt02}) using new high quality Arecibo data. They also used
the measured profile to perform a study of the 2-dimensional beam
pattern of a pulsar for the first time -- a technique that we will
refer to as ``beam tomography'' (see Section \ref{sec:app:beam}).

\subsection{PSR B1534+12}
\label{sec:1534}

The second discovered DNS, PSR B1534+12, has an orbital period of 10 h
and an eccentricity of 0.27. The determined masses result in a
precession rate of 0.5 deg per year, suggesting that the effects of
relativistic spin precession would be measurable. Indeed, this source
was the first one to reveal the predicted changes in polarisation
characteristics, aided by the existence of a highly polarized
interpulse component that enabled a precise RVM fit and the
determination of the impact parameter $\beta$ and its change with time
\cite{arz95,stta00,sta04}.  Moreover, in
addition to the secular profile changes due to spin precession, the
effect of orbital
aberration, causing the cut through the emission beam to change
with orbital phase, was also detected for the first time\cite{sta04},
providing direct evidence that pulsars are indeed {\em
rotating} neutron stars. A combination of these results led to the 
first independent limits on the precession rate that, albeit with low
precision, was consistent with the predicted GR value.

\subsection{PSR J1141$-$6545}
\label{sec:1141}

The binary pulsar J1141$-$6545 is a remarkable system as it harbours a
young 394-ms pulsar in a relativistic eccentric 4.5-h orbit about a
massive white dwarf companion \cite{klm+00a,acw+10}. We therefore observe a
non-recycled pulsar, formed in a recent SN explosion that almost
certainly left a pulsar with a spin-axis that is misaligned with the
orbital momentum vector. With an expected precession rate of 1.4 deg
per year, the pulsar was immediately suggested to be a prime
candidate for spin-precession studies. In fact, spin-precession was
put forward to explain why the rather strong radio source visible as
the young pulsar had not been detected in a previous low-frequency survey
for pulsars, covering the source position in July 1993
\cite{klm+00a}. Unfortunately, like in a cosmic conspiracy, the data for this
particular pointing have been lost, so that this hypothesis cannot be
tested anymore against possible effects of radio interference which may have
masked the pulsar.

The pulsar was monitored independently by two groups, i.e.~our group
around the initial discovery team and the Swinburne group who were the
first to publish the profile variations \cite{hbo05} which are
clearly and easily visible both in total power and polarisation. 
Figure~\ref{fig:1141} presents the results of our monitoring
project that demonstrate that the assumption of a cone-like beam
structure, invoked successfully for modeling PSR B1913+16 (see
Sec.~\ref{sec:1913}), is much too simple for this source. The only way
of obtaining information from the observed relativistic spin
precession for geometry determination, beam tomography or even GR
tests, is therefore given by the attempt to understand the changing
polarisation behaviour of this rather highly polarised source.

The available data indeed offer an opportunity to apply a new method
for studies of relativistic spin precession that was not possible
before. It was noted by Kramer \& Wex (Ref.~\citen{kw09}) that due to the
precession of the spin vector and hence its changing projection on the
sky, the absolute polarisation angle (PA), reflecting this direction,
should also show a periodic variation with time (cf.~Ref.~\citen{lk05}). For
the well-calibrated Parkes data, this absolute position angle
information was available for a sufficiently long time to apply this
method for the first time. Our results published in Ref.~\citen{mks+10}
indeed show the expected PA variation that can be modelled by the
formalism introduced in Ref.~\citen{kw09}. Using a global fit of all
available PA data measured at carefully selected pulse longitudes a
self-consistent description of the geometry and precession parameters
is achieved. In a least-squares fit of only 4 free model parameters
(i.e.~misalignment angle, precession phase, absolute PA offset and
magnetic inclination angle) to 920 data points covering 5 years of
observations, we achieve a reduced $\chi^2=7.4$. We find that the
spin-orbit misalignment angle is about 110 deg.
 At the start of our observations, the impact parameter $\beta$
was about 4 deg in magnitude and it reached a minimum
very close to the magnetic pole around early 2007, consistent with the
observed pulse width variations. We have therefore mapped
approximately one half of the emission beam, which we further discuss
in Sec.~\ref{sec:app:beam}.

PSR J1141$-$6545 was not detected in the Parkes 70cm
survey\cite{mld+96} although, even at the present relatively low flux-density
levels, a detection with signal/noise ratio of the order of 50 would
have been expected. Observations within half a beamwidth of the pulsar
position were made 1993 July 14 (MJD 49182). According to the fitted
geometry model, $\beta$ at that time was about
$−8$ deg. This non-detection therefore suggests that the beam
half-width in latitude is $\sim 8$ deg, although this must be
qualified because of the patchy beam structure.  Observations over the
next decade or two will establish whether or not this is the case for
PSR J1141$-$6545 as $\beta$ returns to large (negative)
values. Indeed, with the reversal in the rate of change of the impact
parameter, we predict that over the next decade we will see a reversed
``replay'' of the variations observed in the past decade, providing
an excellent test for the made modelling. While we will also learn a
big deal about the beam structure, given the fact that the beam does not seem to
follow an organized pattern, it is unlikely that we can use our
observations for a quantitative test of GR.

\begin{figure}[t]
\begin{center}
\psfig{file=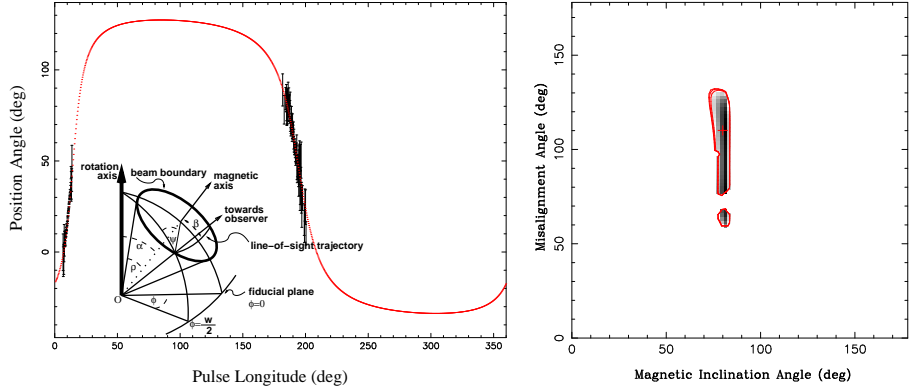,width=12cm} 
\end{center}
\caption{Left) Polarisation angle swing measured for PSR
  J1906+0746. Fitting a simple geometrical ``rotating vector model''
  as shown in the inset, the solid line is derived. Right) Applying
  the
Kramer \& Wex model, the magnetic inclination angle $\alpha$ and
the misalignment angle between pulsar spin and orbital momentum vector
can be constrained (Desvignes et al., in prep).}
\label{fig:1906}
\end{figure}

\subsection{PSR J1906+0746}

This peculiar system shares similarities with the previous system
J1141$-$6545 in that we see the precession of a young, unrecycled
neutron star. In contrast, however, the companion is most likely
another neutron star \cite{lsf+06}.  Due to the young age and its
144-ms period, the timing precision for J1906$+$0746 is unfortunately
limited. However, the orbit is eccentric, $e=0.09$, and rather short,
$P_{\rm b} = 4.1$ h, so that relativistic effects are seen. The
measurement of an periastron advance and an ``Einstein delay''
(i.e.~the combined effect of gravitational redshift and a second-order
Doppler effect) 
imply neutron star masses of $1.35M_\odot$ and $1.26M_\odot$,
respectively, resulting in a GR precession rate of 2.16 deg yr$^{-1}$.
This large precession rate (i.e. twice as large as for the
Hulse-Taylor pulsar and half the size as for the Double Pulsar)
manifests itself easily in the observed pulse profile which (now)
exhibits a strong interpulse separated from the main pulse by half a
period.  The first indication that spin precession is occurring in
this system was indeed noticed when the discovery observation in 2005
was compared with a serendipitous observations from 1998 where the
interpulse was missing \cite{lsf+06}.  In following monitoring
observations it became also clear that the separation of the
interpulse and main pulse is slowly decreasing with time
\cite{kas08,dck+08,des10}, indicating a steady change in
geometry. Fortunately, both main and interpulse of PSR J1906$+$0746
are also highly polarised, so that the RVM fit is very well
constrained due to the wide range of available pulse longitudes during
the fit (see Fig.~\ref{fig:1906}). This allows two ways of measuring the change in geometry:
Firstly, one can fit a RVM separate to data of each epoch where the
uncertainty in derived angles is sufficiently small to detect a systematic
change in the impact angle. Secondly, one can apply the model by
Kramer \& Wex (Ref.~\citen{kw09}) and fit all epochs simultaneously with only
four parameters. With a reduced $\chi^2=1.21$ and 632 degrees of
freedom, the least-squares fit is extremely good, confirming the
suspected orthogonal geometry of the pulsar and a very large
misalignment angle between 60 and 120 deg (see Fig.~\ref{fig:1906}). The possibility to obtain a
precise RVM fit using absolutely calibrated polarisation data promises
a quantitative test of relativistic spin precession in the future.

\subsection{PSR J0737$-$3039 -- The Double Pulsar}
\label{sec:0737}

In 2003 a binary system was discovered where at first one member was
identified as a mildly recycled pulsar with a 23 ms period
\cite{bdp+03} before then the companion was also recognized as a young
radio pulsar with a period of 2.8 s \cite{lbk+04}. Both pulsars, now
known as PSR J0737$-$3039A and PSR J0737$-$3039B, respectively, (or ``A''
and ``B'' hereafter), orbit each other in less than 2.5 hours in a
slightly eccentric orbit. As a result, the system is not only the
first and only double neutron star system where both neutron stars are
visible as active radio pulsars, but it is also the most relativistic
binary pulsar known to date. 

As the most relativistic binary system known to date, we expect a
large amount of spin precession in the Double Pulsar system.  Indeed,
as shown in Tab.~1, the precession periods should be 75 years for A
and 71 years for B. Despite careful studies, profile changes for A
have not been detected, suggesting that A's misalignment angle is
rather small \cite{mkp+05, fsk+08,fer08}. In contrast, changes in the
light curve and pulse shape on secular timescales \cite{bpm+05} reveal
that this is not the case for B. In fact, B had been becoming
progressively weaker and disappeared from our view in 2009
\cite{pmk+10}. Making the valid assumption that this disappearance is
solely caused by relativistic spin precession, it will only be out of
sight temporarily until it reappears later. Modeling suggests that,
depending on the beam shape, this will occur in about 2035 but an
earlier time cannot be excluded. The geometry that is derived from
this modeling is consistent with the results from complementary
observations of spin precession, visible via a rather unexpected
effect described in the following.

The orbit of the Double Pulsar is seen nearly edge on, i.e.~the
inclination angle of the orbit is measured (using a Shapiro delay) to
be 88 deg \cite{ksm+06}.  This leads to $\sim 30$-s long eclipses of A
that are caused by the blocking rotating magnetosphere of B at
superior conjunction. Applying a simple successful geometrical model,
Breton et al. (Ref.~\citen{bkk+08})  were able to explain the regular bursts
of emission of A seen during the dark eclipse phases, which are
separated by a full- or half-period of B. As this pattern is
determined by the three-dimensional orientation of the magnetosphere
of B, which is centred on the precessing pulsar spin, changes in the
eclipse pattern with time were expected and found: eclipse monitoring
over the course of several years shows exactly the expected changes,
with model fitting indicating a constant magnetic inclination angle
and constant misalignment angle, but an azimuthal spin position
changing with a rate of $\Omega_{\rm p, B}= 4.77^{+0.66}_{-0.65}$ deg yr$^{-1}$.
As shown in Tab.~1, this value is fully consistent with the value
expected GR. This measurement, however, also allows to tests alternative
theories of gravity and their prediction for relativistic
spin-precession in strongly self-gravitating bodies for the first time (see
Section~\ref{sec:app:gr}).

\begin{figure}[t]
\begin{center}
\psfig{file=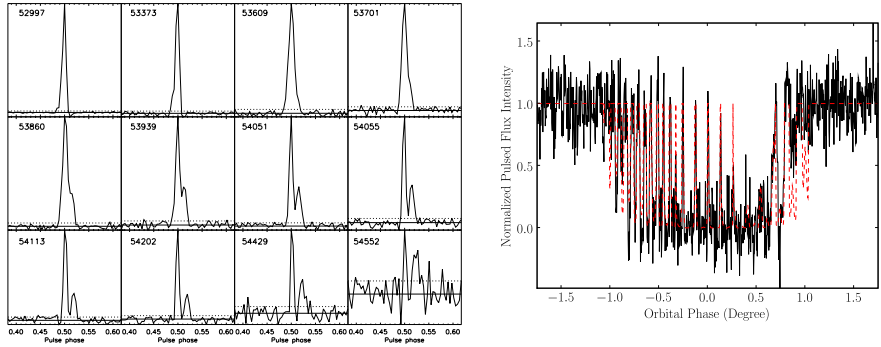,width=12cm} 
\end{center}
\caption{Left) Pulse profiles of B in the Double Pulsar at 12
  different days demonstrating the disappearance of the
  pulsar\cite{pmk+10}. Right) Observed eclipse of pulsar A caused by
  the magnetosphere of B. The distinctive pattern depends on B's
  orientation and changes as a function of time\cite{bkk+08}.}
\label{fig:0737}
\end{figure}

\section{Applications}

As demonstrated, relativistic spin precession is now observed in a variety of sources
and has become by now a tool with applications that go beyond simply
detecting the effect as a further phenomenon predicted by GR. Firstly,
however, we discuss exactly this usage of spin-precession.

\subsection{Testing theories of gravity}
\label{sec:app:gr}

The phenomena observed in the pulse structure of relativistic pulsars
is consistent with the predictions of relativistic spin
precession. For a real test of GR or other alternative theories of
gravity a quantitative measurement of spin precession parameters are
however needed. The simplest approach is to measure the rate of
precession, $\Omega_p$, and to compare it with the expected values
(Tab.~1). Usually, such a measurement is difficult, as the shape and
structure of the pulsar beam is unknown but often needed in modeling
geodetic precession. In fact, spin precession is often used (as the
only way!)  to {\em infer} 2-D beam information (see
Section~\ref{sec:app:beam}. This is different when polarisation
measurements reveal the (changing) geometry. Indeed, the first
quantitative measurement of relativistic spin-precession in binary
pulsars was possible with polarisation measurements of PSR B1534+12
(see Section~\ref{sec:1534}) albeit with limited precision. A much
better test was enabled with the spin precession seen via the eclipses
in the Double Pulsar (see Section~\ref{sec:0737} where a 13\%
precision was achieved in a strong-field regime. Due to our ability
to independently measure both orbits in the system, we can use the
result also to put first constraints on alternative theories of
gravity.

After the introducing of the method by Breton et al.~(Ref.~\citen{bkk+08}),
Kramer \& Wex (Ref.~\citen{kw09}) describe in detail how the relativistic
spin-precession rate of pulsar B (see Section \ref{sec:0737}), can be interpreted in
formalism introduced by Will\cite{wil93} and Damour and Taylor
\cite{dt92}. These authors constructed a Lagrangian that generalizes
the Lagrangian of the post-Newtonian orbital dynamics in a
strong-field regime, for fully conservative theories of gravity
(modified Einstein–Infeld–Hoffmann formalism).  In particular, to
account for strong-field effects in the spin–orbit interaction, Damour
\& Taylor introduced the coupling function $\Gamma_i^j$ in the
spin-orbit Lagrangian of Ref.~\citen{dam82}, where the indices $i$ and
$j$ refer to the two bodies in the system. The measured relativistic
spin-precession rate of pulsar B can now be used to limit, for the
first time, $\Gamma_{\rm B}^A$ as
\begin{equation}
  \Omega_{\rm p, B} = n_{\rm b}X_{\rm A}X_{\rm B}\left[(1+R)
                            \frac{\Gamma_{\rm B}^{\rm A}}{\cal G} - \frac{R}{2}
                            \right]\frac{\beta_O^2}{1-e^2} \:,
\end{equation}
where $\beta_O \equiv({\cal G}Mn_{\rm b})^{1/3}/c$ is a characteristic
velocity for the relative orbital motion and $n_{\rm b} = 2\pi/P_{\rm
  b}$ is the orbital frequency.  In general relativity ${\cal G} =
G$, but in alternative theories of gravity the actual value depends on
the parameters of the theory and of the structure of each body. In
other words, for neutron stars these parameters can deviate
significantly from their values in general relativity, even if their
weak-field limit agrees with general relativity \cite{wil93,de96}.
The fact that the Double Pulsar gives access to the mass ratio,
\begin{equation}
R \equiv \frac{m_{\rm{A}}}{m_{\rm{B}}} 
  =      \frac{a_{\rm{B}}}{a_{\rm{A}}} 
  =      \frac{a_{\rm{B}}\sin i/c}{a_{\rm{A}}\sin i/c} 
  \equiv \frac{x_{\rm{B}}}{x_{\rm{A}}} \:.
\end{equation}
in any
Lorentz-invariant theory of gravity \cite{lbk+04,dam07}, allows us to determine $X_{\rm A} = R/(1+R)
= 0.51724 \pm 0.00026$ and $X_{\rm B}=1/(1+R) = 0.48276 \pm
0.00026$. As detailed in Ref.~\citen{kw09}, with this
information the measurement of the shape of the Shapiro delay, $s$, can be
used to determine $\beta_O $ via
\begin{equation}
  s          = \frac{n_{\rm b}x_{\rm A}}{\beta_O X_{\rm B}} \:,
\end{equation}
to $\beta_O = (2.0854 \pm
0.0014) \times 10^{-3}$. 
Consequently
\begin{equation}
  \frac{\Gamma_{\rm B}^{\rm A}}{2\cal G} = 0.95 \pm 0.11 \:,
\end{equation}
As pointed out in Ref.~\citen{kw09}.  this is not only in agreement with
general relativity, which predicts $\Gamma_{\rm B}^{\rm A}/2{\cal G} =
1$, but it also demonstrates that the relativistic precession of a
spinning body is independent of its internal structure. They emphasise
that currently the Double Pulsar is the only system that allows for
the test of the ``effacement'' property of a spinning body.

\subsection{Core collapse physics \& neutron star birth}
\label{sec:app:sn}

As described in Section~\ref{sec:evolv}, the evolutionary history of
most binary pulsars involves a phase of mass transfer after which we
would usually expect the spin vectors of the stars to be aligned with
the orbital momentum vector. If a supernova (SN) occurs after this
alignment, a kick may be imparted on the newly born neutron star,
tilting the post-SN orbit relative to the pre-SN configuration.
Modeling the observed effects of relativistic spin precession, we can
determine the misalignment angle between the spin vector of the
precessing neutron star (see Ref.~\citen{kra98} and Sections above). This
information can be used to infer the pre-SN configuration which can
then be compared to the observed post-SN situation in order to learn
about the SN explosion itself, such as the kick amplitude and
direction.

Such an application of spin precession was first achieved by Wex et
al.~(Ref.~\citen{wkk00}) who used the geometry of the PSR B1913+16
system \cite{kra98} to derive that the kick was relatively large and
its direction rather well confined.  This type of analysis involves
tracing the motion of the binary back in time to potential birthplaces
in the plane of the galaxy, assuming that the large binary mass
remaining after the first SN meant that the space velocity at that
time would not be large. The gravitational radiation decay of the
orbital eccentricity and separation must also be corrected for
\cite{pet64}. The kick magnitude and progenitor mass can then be
constrained using simple equations \cite{kal96}.  To achieve the
result, one also has to make assumptions about the possible
characteristics of the exploding He-star. The latter input and the
applied techniques were subsequently refined and applied by Kalogera
and her group (e.g.~Refs.~\citen{ wkh05,kvm08}) who performed similar work
for instance for PSR B1534+12 and the Double Pulsar. Their results can
be compared with work by other authors, who all agree that spin
precession results for PSR B1534+12 suggest that its companion
received a large kick with velocities of $200-270$ km s$^{-1}$ and
progenitor masses were of $2.00-3.35 M_\odot$ \cite{tds05,
  wkh05}. For the Double Pulsar the results of similar
calculations differ, partly because of the role of the unknown radial
velocity of the system, but also because the system has a number of
peculiar properties.

On one hand, the mass of pulsar B is small, only $1.25M_\odot$
\cite{ksm+06}, prompting the argument that B might not have formed in
an iron core collapse but may have been born in an electron-capture
supernova, in which a slightly less massive O-Ne-Mg core captures
electrons onto Mg to initiate the collapse \cite{pdl+05}. On the other
hand, the derived very small systemic velocity of the Double Pulsar
(corrected for Galactic rotatation) of only $9^{+6}_{-3}$ km s$^{-1}$
(see Refs.~\citen{ksm+06,dbt09}) indicates that, depending on the progenitor mass
of B, the kick velocity was small (e.g.~Refs.~\citen{ps05,ps06,wkf+06,std+06})
but see the review by Kramer \& Stairs (Ref.~\citen{ks08}) for a detailed
discussion of the various results. What seems to be clear from the absence of
any profile changes in pulsar A, is that the misalignment vector in A must be rather small,
consistent with a ``gentle'', low-velocity kick birth of young pulsar
B (e.g.~Ref.~\citen{std+06}).

Finally, a similar study based on the geometry derived from spin
precession for PSR J1141$-$6545 suggests that the pre-supernova star had
a mass of only $\sim2 M_\odot$ and that the supernova kick velocity
was relatively small, between 100 and 250 km s$^{-1}$ depending on the
assumed systemic velocity \cite{mks+10}.

\begin{figure}[t]
\begin{center}
\psfig{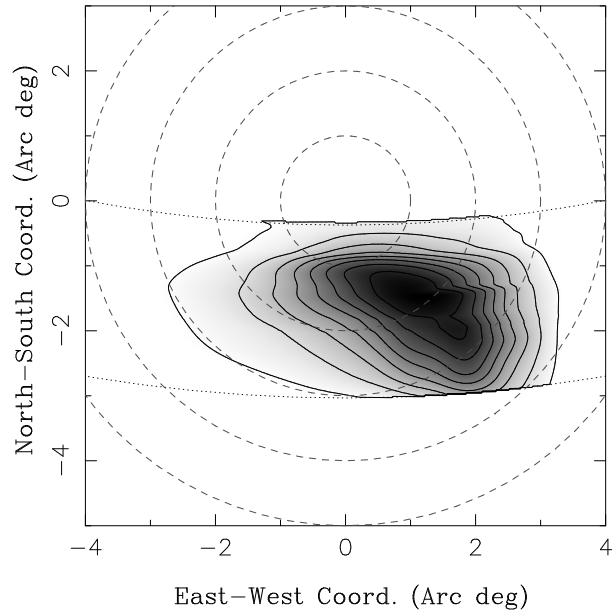} 
\end{center}
\caption{Beam map derived for PSR J1141$-$6545 in Ref.~\citen{mks+10}.}
\label{fig:1141beam}
\end{figure}

\subsection{Pulsar Beam Tomography}

\label{sec:app:beam}

As our changing line-of-sight intersects the pulsar emission beam,
relativistic spin precession provides the only way to actually map the
structure of the radio beam. This idea \cite{dr74,wrt89} of a ``beam
tomography'' was first beautifully applied to PSR B1913+16
\cite{wt00,wt02,cw08} where it was shown that the simple hollow cone
structure (filled with a core component) is perhaps oversimplified
but not too far from the truth. Deviations from a purely circular beam
are observed, suggesting an elongation of the beam in North-South
direction. The initally suggested hourglass shape of the emission
beam \cite{wt00,wt02} appears to be too extreme given the new improved
modeling \cite{cw08}.

It is interesting to see whether the emission beams of recycled
pulsars look different from those of young pulsars. The first
non-recycled pulsars for which we can infer the beam pattern is
PSR J1141$-$6545 where the geometry suggests that we
only see one half of the polar-cap region. Despite this we can clearly
see that the beam is quite asymmetric with no evidence for a core-cone
or ring structure that is symmetric about the magnetic pole. The
partially filled beam can be described as ``patchy'', albeit with just
one major patch in the region scanned so far. This is the
first two-dimensional map of an emission beam to clearly show such
patchy structure (Fig.~\ref{fig:1141beam}). Even though the observed pulse {\em width} is about
average,  the inferred beam {\em radius} is rather small, given that
the so far observed emitting region fits within a circle
of radius about 4 deg centered on the magnetic axis. This is much
smaller than the $\sim 10$ deg expected from the period scaling
derived from other pulsars (e.g.~Refs.~\cite{gks93,gou94,kwj+94}). Time will tell --
by transversing more of the pulsar beam -- whether the beam really
happens to be smaller than usual or whether a large elongation of the
pulsar beam is present.

We already indicated that beam tomography can also be performed for
pulsar B in the Double Pulsar system. Here, the results show that
the observed beam structure cannot be easily explained with a
circular hollow-cone beam either, but that the beam appears to be
elliptical and horse-shoe shaped. This unusual shape may find its
origin in the strong interaction of the pulsar wind of A with the
magentosphere of B, which distorts B's magnetosphere and produces a
cometary shape. As a result, B was only strongly detectable in two
specific orbital phase ranges \cite{lbk+04}. We can therefore expect
the magentospheric currents to be significantly different from
``normal'', undisturbed magnetsopheres, so that the result is extremely
interesting in order to learn more about the exotic conditions in this
system, but it is less likely that the derived beam tomography gives a
representative example for the whole pulsar population.

\section{Summary \& Conclusions}

From the humble beginnings of detecting profile changes in the
Hulse-Taylor pulsar as the first observational evidence for
relativistic spin precssion in binary pulsars, the study of the
associated effects has become routine. Indeed, it is fair to say that
relativistic spin precession has been established as a tool for the
study of physical and astrophyical problems. Today, studies of spin
precession have led to the first and only constraints for spin-orbit
coupling of  strongly self-gravitating bodies, it has been used to reveal
the previously unknown structure of pulsar emission beams and it has
been instrumental in providing evidence that the magnitude of kicks
imparted on neutron stars during their birth can cover a wide range of
magnitude, essentially from 10 to 1000 km s$^{-1}$. In the years to
come, further studies will reveal new results and we indeed look
forward to the observations of spin precession in the first pulsar -
black hole system, the discovery of which we await so eagerly.

\section*{Acknowledgments}

I am grateful to the conference organisers for their hospitality 
and a memorable event.


\end{document}